\documentclass[useAMS,referee,usenatbib]{biom}

\usepackage{float}
\usepackage{pgf}
\usepackage{tikz}
\usetikzlibrary{positioning}
\usepackage{graphicx}
\usepackage{caption}
\usepackage{subcaption}
\usepackage{lscape}
\usepackage{multirow}
\usepackage[colorlinks,bookmarksopen,bookmarksnumbered,citecolor=blue,urlcolor=blue]{hyperref}
\usepackage{amsmath}

 \usepackage{colortbl}

\def\bSig\boldsymbol{\Sigma}

\title[Marginal treatment effect in CRTs with missing outcomes]{Accounting for interactions and complex inter-subject dependency in estimating treatment effect in cluster randomized trials with missing outcomes}

\author{Melanie Prague$^{1,*}$\email{mprague@hsph.harvard.edu}, 
Rui Wang$^{1,2}$, Alisa Stephens$^{3}$, \textbf{Eric Tchetgen Tchetgen$^{4}$ and Victor DeGruttola$^{1}$} \\
\vspace{-0.25cm}$^{1}$ Department of Biostatistics, Harvard T.H. Chan School of Public Health, Boston, MA, U.S.A. \\
\vspace{-0.25cm}
$^{2}$ Division of Sleep and Circadian Disorders, Departments of Medicine and Neurology,\\\vspace{-0.25cm} Brigham and Women's Hospital, Boston, MA, U.S.A. \\
\vspace{-0.25cm}$^{3}$ Center for Clinical Epidemiology and Biostatistics, Perelman School of Medicine, \\\vspace{-0.25cm} University of Pennsylvania, Philadelphia, PA, U.S.A. \\
\vspace{-0.25cm}$^{4}$ Department of Epidemiology, Harvard T.H. Chan School of Public Health, Boston, MA, U.S.A.}

\begin{document}

\date{{\it Received February} 2015. {\it Revised ???} ???.  {\it
Accepted ???} ???.}

\pagerange{\pageref{firstpage}--\pageref{lastpage}} 
\volume{64}
\pubyear{2015}
\artmonth{December}

\doi{10.1111/j.1541-0420.2005.00454.x}

\label{firstpage}

\begin{abstract}
Semi-parametric methods are often used for the estimation of intervention effects on correlated outcomes in cluster-randomized trials (CRTs). When outcomes are missing at random (MAR), Inverse Probability Weighted (IPW) methods incorporating baseline covariates can be used to deal with informative missingness. Also, augmented generalized estimating equations (AUG) correct for imbalance in baseline covariates but need to be extended for MAR outcomes.  However, in the presence of interactions between treatment and baseline covariates, neither method alone produces consistent estimates for the marginal treatment effect if the model for interaction is not correctly specified. We propose an AUG-IPW estimator that weights by the inverse of the probability of being a complete case and allows different outcome models in each intervention arm. This estimator is doubly robust (DR), it gives correct estimates whether the missing data process or the outcome model is correctly specified. We consider the problem of covariate interference which arises when the outcome of an individual may depend on covariates of other individuals. When interfering covariates are not modeled, the DR property prevents bias as long as covariate interference is not present simultaneously for the outcome and the missingness. An R package is developed implementing the proposed method. An extensive simulation study and an application to a CRT of HIV risk reduction-intervention in South Africa illustrate the method. \\
 
\end{abstract}

\begin{keywords}
Augmentation; Cluster-randomized trials; Generalized estimating equation (GEE); Interactions; Interference; Inverse probability weighting (IPW); Missing at random (MAR); Outcome Model; Propensity Score; R package; Semi-parametric methods.
\end{keywords}

\maketitle

\section{Introduction}

 In clustered randomized clinical trials (CRTs), the unit of treatment assignment is a cluster of subjects, which we also refer to as community.  In such settings, outcomes are likely to be correlated among subjects within the same cluster. Often used for estimation, generalized estimating equations (GEE) based on semi-parametric methods \citep{zeger1986longitudinal} target marginal effects of treatment. Within clusters, dependence can be modeled using a working correlation structure. Compared to mixed effects models, this approach has the advantage of focusing on population average effects rather than  cluster specific effects (which are equal for continuous outcomes) and requires fewer parametric assumptions on the outcome distribution \citep{hubbard2010gee}. \textcolor{black}{Moreover, because both the outcome and the missing data mechanism can be modeled, this approach allows doubly robust estimation, which is impossible with mixed effect models.} Finally, this approach to estimation is robust to misspecification of the correlation structure. However, challenges arise in developing a consistent and efficient estimator of  marginal treatment effects; these include the need to adjust for missing data and accommodate covariate interference (wherein a subject's outcome may be affected by covariates of other subjects) and interactions (wherein the effect of treatment varies by covariate-defined subgroups). We propose a method that addresses these issues and is practical to implement for evaluating novel interventions in CRTs.

 In CRTs,  covariates may be fully observed even if the outcome is missing. When data are assumed missing completely at random (MCAR) -- i.e. the observed process is independent of observed and unobserved information \citep{rubin1976inference} -- the standard GEE approach provides consistent and asymptotically normal (CAN) estimators. If the pattern of missingness depends on observed information but not on missing data, the data are said to be Missing at Random (MAR). In this setting,  the standard GEE may yield biased estimates although likelihood-based approaches, such as mixed effect models, can provide unbiased estimators. Imputation \citep{paik1997generalized} or reweighing  \citep{robins1995analysis} methods can correct for this bias. Although useful if the missingness mechanism is not completely known, multiple imputation requires correct specification of the joint distribution of the outcomes, which is especially difficult when they are correlated and the cluster sizes are large \citep{beunckens2008simulation}. In this article, we consider the Inverse Probability Weighting  (IPW) approach to analyze incomplete data. If the model for the missingness mechanism represents  the MAR data generating process, the IPW estimation provides CAN estimators of treatment effects by  reweighing complete cases according to the probability of being observed \citep{liang1986longitudinal,robins1994estimation}. 

Recent methodological developments  improve estimation efficiency by leveraging baseline covariates; they may be based on targeted maximum likelihood  \citep{moore2009increasing} and on augmentation \citep{robins1994estimation,robins2000marginal,tsiatis2008covariate,zhang2008improving}.~\citet{stephens2012augmented} developed  the augmented GEE (AUG) methods in the setting of dependent outcomes such as in CRTs. The AUG adds a  term to the standard GEE which relates the outcome to covariates and treatment. Randomization assures that the AUG is CAN even in the case of OM misspecification.  However in the case of outcome data that are MAR but not MCAR, the AUG may be biased.  There exists theory for extending these methods to  MAR data  for individual randomized Trials (RTs) with possibly correlated data \citep{van2003unified,glynn2010introduction}, we focus on the details of implementing the methods in CRTs.  
 
The term interference can refer to different types of relationships among exposures, outcomes and covariates. 
Interference in RTs arises when one subject's treatment may impact the outcomes of other subjects  \citep{rosenbaum2007interference,vansteelandt2007confounding,tchetgen2012causal,hudgens2012toward}. A similar phenomenon, confounding by clusters, has been discussed in the context of observational studies \citep{seaman2014review}; we will refer to such confounding as exposure interference. In CRTs all subjects within a cluster receive the same treatment; hence if the clusters are independent as typically assumed in practice, there is no exposure interference measured at the cluster level. Therefore, any choice of working correlation structure for the standard GEE will give a consistent estimator of the marginal treatment effect \citep{sullivan1994cautionary}. We will investigate covariate interference among individuals nested within clusters: the setting in which one subject's covariate may impact the outcomes of other subjects.

 The IPW and the AUG can be combined in a doubly-robust method we refer to as the DR; we investigate its properties regarding robustness to misspecification of the missing data and outcome generating process. By considering a variety of data generating mechanisms, we investigate settings in which the DR has advantageous properties (consistency and precision) compared to the IPW and the AUG, and discuss the impact of covariate interference and treatment-covariate interactions. This paper is organized as follows. Section 2 introduces notation and assumptions for the IPW and the AUG GEE approaches. Section 3 describes the DR approach, investigates CAN properties and discusses the issue of covariate interference. Section 4 provides a motivating example with data arising from a CRT of an HIV / Sexually Transmitted Infection (STI) risk reduction intervention in South Africa  \citep{jemmott2014cluster}. Simulation studies regarding bias, relative efficiency and coverage are described in Section 5, and concluding remarks are made in Section 6.

\section{Notation, basic models and assumptions}
\label{part:notation}
\subsection{Notation for CRTs and marginal treatment effect}
We consider a study design in which a vector of $P$ baseline covariates $\boldsymbol X_{ij}=(X^1_{ij}, \dots, X^P_{ij})$ and outcome $ Y_{ij}$ are recorded for each subject
$j=1,\dots,  n_{i}$ in community $i=1 ,\dots,  M$. The sample size within each community is assumed fixed by design and non-informative. Our setting compares two arms (treated $A_{i}=1$ and control $ A_{i}= 0$); the probability of treatment assignment is known and given by $p=P(A_i=1)$; extension to a greater number of treatments is straightforward but complicates the  notation.  In this article, the outcome $\boldsymbol Y_i=[Y_{ij}]_{j=1,\dots,  n_i}$ is assumed to be continuous, but extension to other types of outcomes is straightforward. The vector $\boldsymbol{R}_i=[R_{ij}]_{j=1,\dots,  n_i}$ is the indicator of missingness;  $Y_{ij}$ is observed when $R_{ij}=1$. The matrix of covariates $\boldsymbol{X}_i=[\boldsymbol X_{ij}]_{j=1,\dots,  n_i}$ is assumed to be fully observed and consists only of pre-exposure covariates measured at baseline. 

Interest lies in estimating the marginal effect of the treatment given by $ M_E^*=E(E( Y_{ij}|A_i=1, \boldsymbol X_i)-E( Y_{ij}| A_i=0, \boldsymbol X_i))$.   For estimating $M_E^*$, we  make inference about the parameters $\boldsymbol \beta = (\beta_0,\beta_A)^T$ indexing the marginal model $g(\mu_{ij}(\boldsymbol \beta, A_i))=g(E(Y_{ij}|A_i))=\beta_0+\beta_A A_{i}$, where $\boldsymbol \mu_i(\boldsymbol \beta, A_i)=[\mu_{ij}(\boldsymbol \beta, A_i)]_{j=1,\dots,n_i}$ and $g$ is a one-to-one link function, which is an identity function in this article. Of particular interest, $\beta_A$ is equal to $M_E^*$. Of note, extension to binary outcome $Y_{ij}$ using a logistic function for $g$ and considering odd-ratios is based on the same reasoning.

 When the outcome is  believed to be MCAR, the missingness process is independent of $\boldsymbol X_i$, $A_i$, and $\boldsymbol Y_i$. If one assumes MAR and the missingness pattern is monotone, the probability of missingness can be estimated by a multistep approach by decomposing a monotone missing pattern into multiple uniform missing data models \citep{robins1994estimation,li2011weighting}. In CRTs, any component of $\boldsymbol{Y}_i$ can be missing; hence the missingness pattern is  non-monotone. Therefore, we make a stronger assumption than MAR that we refer to as restricted MAR (rMAR): the probability that the outcome  for one individual is missing is independent of all outcomes in the cluster, conditional on baseline exposure $A_i$ and cluster characteristics $\boldsymbol X_i$. The conditional probability that the outcome is observed is denoted $\pi_{ij}(\boldsymbol X_i, A_i)={P}(R_{ij}=1|\boldsymbol{X}_i,A_i)$ and is called the propensity score (PS). When data are rMAR, ignoring missing data leads to biased inference if missingness depends both on $\boldsymbol X_i$ and $A_i$.  This is because the presence of missing data no longer assures balance on average of confounding factors between treatment arms.  Therefore, analysis must include adjustment for missing data; appropriate models for this adjustment may require treatment-covariate interactions, which may be difficult to specify and require many parameters. Combining the IPW and the AUG, which this paper proposes, makes it possible to obtain  consistent estimates of the marginal effect of treatment without explicitly specifying interaction terms while also improving efficiency. 

\subsection{Inverse Probability Weighted Generalized Estimating Equations (IPW) }
\label{part:IPWGEE}
In order to account for missing data, semi-parametric estimators based on the IPW are found by solving the estimating equation \ref{eq:IPWGEE}:
\begin{eqnarray} 
0 &=&\sum_{i=1}^M \underbrace{\boldsymbol D_i^T \boldsymbol  V_i^{-1} \boldsymbol  W_i(\boldsymbol X_i, A_i, \boldsymbol \eta_W) \left[ \boldsymbol  Y_i -\boldsymbol \mu_i(\boldsymbol \beta,A_i) \right]}_{\boldsymbol\psi_i(\boldsymbol Y_i, \boldsymbol R_i,  A_i, \boldsymbol \beta, \boldsymbol \eta_W)},
\label{eq:IPWGEE} 
\end{eqnarray} 
\normalsize
 where $\boldsymbol D_i=\frac{\partial \boldsymbol \mu_i(\boldsymbol \beta,A_i)}{\partial \boldsymbol \beta^T}$ is the design matrix, $\boldsymbol V_i$ is the covariance matrix equal to $\boldsymbol U_i^{1/2} \boldsymbol C(\boldsymbol \alpha)\boldsymbol U_i^{1/2}$ with $\boldsymbol U_i$ a diagonal matrix with elements ${\rm var}(y_{ij})$ and $\boldsymbol C(\boldsymbol \alpha)$ is the working correlation structure with non-diagonal terms $\boldsymbol \alpha$. For example, for an independence correlation structure $\boldsymbol \alpha$ are zero; for exchangeable all the elements of $\boldsymbol \alpha$ are identical. Parameters $\boldsymbol \alpha$ could also depend on the treatment group $\boldsymbol C(\boldsymbol \alpha(A_i))$ but we do not consider this possibility in our implementation. In this article, we estimate the $\boldsymbol \alpha$ parameters using moment estimators from the Pearson residuals as in \citet{mcdaniel2013fast}. The $n_i \times n_i$ matrix of weights is $\boldsymbol W_i(\boldsymbol X_i, A_i, \boldsymbol \eta_W)=diag\left[R_{ij}/\pi_{ij}(\boldsymbol X_i, A_i, \boldsymbol \eta_W)\right]_{j=1,\dots,n_{i}}$, where the PS is derived by fitting a binary response model to the indicator $R_{ij}$ regressed on $A_i$ and a subset of $\boldsymbol X_{i}$ -- say using a logistic regression. The $\boldsymbol \eta_W$ are nuisance parameters estimated in the PS. A necessary assumption for this method is that  probabilities for the PS are  bounded away from zero. Several authors have noted the instability that may arise from small probabilities of observation (i.e. large weights) and proposed use of stabilized or truncated weights; see \citet{seaman2013review} for a review. To ensure that the IPW provides a CAN estimator, the PS must include all covariates that are associated simultaneously with the missingness and outcome \citep{brookhart2006variable}, including those that involve interaction with treatment \citep{belitser2011measuring}.

\subsection{Augmented Generalized Estimating Equations (AUG) }
\label{part:AugGEE}
For settings with complete data, \citet{stephens2012augmented} proposed the AUG estimator  which can  improve efficiency relative to the standard GEE by incorporating baseline covariates. The AUG is constructed by subtracting from the set of GEEs the orthogonal projection of the standard estimating function onto the span of scores corresponding to all smooth parametric models for the treatment assignment mechanism given covariates. The AUG is given in Equation \ref{eq:AGEE}: 
\begin{eqnarray} 
0&=& \sum_{i=1}^M \Bigg[ \underbrace{\boldsymbol D_i^T \boldsymbol V_i^{-1} \left( \boldsymbol Y_i -\boldsymbol \mu_i(\boldsymbol \beta,A_i)\right)}_{\boldsymbol{\tilde{\psi}}_i(\boldsymbol Y_i,A_i,\boldsymbol \beta)} \nonumber \\
& & \quad + \sum_{a=0,1} p^a(1-p)^{1-a} \boldsymbol D_i^T \boldsymbol V_i^{-1}  \Big( \boldsymbol B(\boldsymbol X_i,A_i=a,\boldsymbol \eta_B) -\boldsymbol \mu_i(\boldsymbol \beta,A_i=a)\Big) \Bigg] .
\label{eq:AGEE} 
\end{eqnarray} 
\normalsize 

The term $\boldsymbol{\tilde{\psi}}_i(\boldsymbol Y_i, A_i, \boldsymbol \beta)$ is similar to $\boldsymbol \psi_i(\boldsymbol Y_i, \boldsymbol R_i,   A_i, \boldsymbol \beta, \boldsymbol \eta_W)$ in Equation \ref{eq:IPWGEE} for the IPW except that $\boldsymbol W_i$ is set to identity because there is no adjustment for missing data. Definitions for $\boldsymbol D_i$ and $\boldsymbol V_i$ remain the same. \textcolor{black}{The function $\boldsymbol B(\boldsymbol X_i,A_i=a,\boldsymbol \eta_B)$ is an arbitrary function of $\boldsymbol X_i$ given for each treatment arm.  The $\boldsymbol \eta_B$ are nuisance parameters that must be estimated. The estimator in Equation \ref{eq:AGEE} is most efficient if $\boldsymbol B(\boldsymbol X_i,A_i=a,\boldsymbol \eta_B)$ models the outcome and is equal to $E( Y_{ij}|\boldsymbol X_i,A_i=a)$ \citep{robins1994estimation,zhang2008improving}. Thus, $\boldsymbol B(\boldsymbol X_i,A_i=a,\boldsymbol \eta_B)$ is called the outcome model (OM).} In the absence of missing data, the AUG remains consistent even if the OM is not correctly specified ($\boldsymbol B(\boldsymbol X_i,A_i=a,\boldsymbol \eta_B)\neq E( Y_{ij}|\boldsymbol X_i,A_i=a)$).  Correct specification can lead to substantial efficiency gains compared to the standard GEE. Moreover, in presence of treatment-covariate interactions, it is useful to fit a different regression model for the OM for each treatment group, e.g. $\boldsymbol B(\boldsymbol X_i,A_i=a,\boldsymbol \eta_B)=\gamma^a_0+ \sum_{r=1}^P\gamma^a_r X^r_{ij}$ with $\boldsymbol \eta_B=(\gamma^0_1, \dots, \gamma^0_P,\gamma^1_1, \dots, \gamma^1_P)$, thereby obviating the need to fit covariate-treatment interactions terms. In presence of rMAR, the AUG does not ensure consistent estimation; instead, one must combine the AUG with the IPW as we show below.

\section{Methods to accommodate missing data, treatment-covariate interactions and covariate interference in CRTs}
\subsection{Doubly Robust Augmented IPW Generalized Estimating Equations (DR)} 
\label{part:supp1}

We  extend the AUG in Equation \ref{eq:AGEE} to account for missing data using the IPW in Equation \ref{eq:IPWGEE} by subtracting from the set of GEEs the orthogonal projection of $\boldsymbol\psi_i(\boldsymbol  Y_i, \boldsymbol  R_i,   A_i,\boldsymbol \beta, \boldsymbol \eta_W)$ onto the span of scores corresponding to all smooth parametric models for the missing data process and the treatment assignment mechanism  given covariates \citep{tsiatis2006improving}. This gives the following estimating equation (see Web-Supplementary Material B for details):
\begin{eqnarray}
0&=& \sum_{i=1}^M \Bigg[ \boldsymbol D_i^T \boldsymbol V_i^{-1} \boldsymbol  W_i(\boldsymbol X_i, A_i, \boldsymbol \eta_W) \left( \boldsymbol Y_i - \boldsymbol B(\boldsymbol X_i, A_i, \boldsymbol \eta_B) \right)\nonumber \\
& &  \quad+ \sum_{a=0,1} p^a(1-p)^{1-a} \boldsymbol D_i^T \boldsymbol V_i^{-1}  \Big( \boldsymbol B(\boldsymbol X_i,A_i=a, \boldsymbol \eta_B) -\boldsymbol \mu_i(\boldsymbol \beta,A_i=a)\Big) \Bigg], \label{eq:DRAIPW} \\
&=& \sum_{i=1}^M \boldsymbol \Phi_i(\boldsymbol Y_i, \boldsymbol R_i, A_i, \boldsymbol X_i, \boldsymbol \beta, \boldsymbol \eta_W, \boldsymbol \eta_B). \nonumber
\end{eqnarray}

 The $\boldsymbol D_i$, $\boldsymbol V_i$ and the PS are defined such as in Equation \ref{eq:IPWGEE}, the OM denoted $ \boldsymbol B(\boldsymbol X_i,A_i=a, \boldsymbol \eta_B)$ is defined for each treatment group such as in Equation \ref{eq:AGEE}. The estimator denoted $\hat{\beta}_{aug}$ is found by solving the estimating equation given in equation \ref{eq:DRAIPW}. Although  analytic solutions sometimes exist, coefficient estimates are generally obtained using an iterative procedure such as the Newton-Raphson method. To get $\hat{\boldsymbol \beta}_{aug} $ we use the estimated PS ($\boldsymbol W_i(\boldsymbol X_i, A_i, \hat{\boldsymbol \eta}_W)$) and estimated OM ($\boldsymbol B(\boldsymbol X_i, A_i, \hat{\boldsymbol \eta}_B)$). As mentioned above, treatment-covariate interactions can be accounted for by fitting  OM regressions separately by treatment group. One could also estimate parameters of  the PS model separately by treatment groups. This approach, however,  may provide less stable results due to variability in the calculation of weights. In this paper, $\hat{ \boldsymbol \eta}_W$ in $\boldsymbol W_i(\boldsymbol X_i, A_i, \hat{ \boldsymbol \eta}_W)$ are obtained using a logistic regression and $\hat{ \boldsymbol \eta}_B$ in $\boldsymbol B(\boldsymbol X_i,A_i,\hat{ \boldsymbol \eta}_B)$ are obtained using a linear regression. Thus, we treat $R_{ij}$ and $R_{ij'}$ as conditionally independent given $A_i$ and $\boldsymbol X_i$. In the presence of correlation of $R_{ij}$ and $R_{ij'}$ , one might be able to improve efficiency of estimation of $\pi_{ij}$ and therefore of the marginal treatment effect by accounting for this correlation. Of note, estimation procedures other than generalized linear models could also be used to compute the OM and the PS values. The DR estimator is doubly robust in the sense that it is CAN under correct specification of  either the OM (i.e. $\boldsymbol B(\boldsymbol X_i,A_i,\hat{ \boldsymbol \eta}_B)=E( Y_{ij} | A_i,\boldsymbol X_i)$) or the PS  (i.e. $\pi_{ij}(\boldsymbol X_i, A_i, \hat{ \boldsymbol \eta}_W)={P}(R_{ij}=1|\boldsymbol{X}_i, {A_i})$) (see Web-Supplementary Material Section C1). Implementation in R is available on the CRAN in the package \textit{'CRTgeeDR'}. Source code had been made available as Web-Supplementary material. We note that in contrast with several existing software packages (for example proc GENMOD in \cite{SASGENMOD}), our implementation of the weighted GEE, which uses $ \boldsymbol V_i^{-1} \boldsymbol  W_i(\boldsymbol X_i, A_i, \boldsymbol \eta_W)$ instead of $ \boldsymbol W^{1/2}_i(\boldsymbol X_i, A_i, \boldsymbol \eta_W) \boldsymbol V_i^{-1} \boldsymbol  W^{1/2}_i(\boldsymbol X_i, A_i, \boldsymbol \eta_W)$, guarantees consistency for all choices of working correlation structure (see details in  Web-Supplementary Material Section C2 and D). 

\vspace{-0.5cm}
\subsection{Variance of the DR estimator}  
The variance of $\hat{\boldsymbol \beta}_{aug}$ is estimated by the sandwich variance estimator. There are two external sources of variability that need to be accounted for: estimation of $\boldsymbol \eta_W$ for the PS and of $\boldsymbol \eta_B$ for the OM. We denote $\boldsymbol \Omega=(\boldsymbol \beta, \boldsymbol \eta_W,\boldsymbol \eta_B)$ the estimated parameters of interest and nuisance parameters. We can stack estimating functions and score functions for $\boldsymbol \Omega$:
$$\small \boldsymbol U_i(\boldsymbol \Omega)= \left( \begin{array}{c}
  \boldsymbol \Phi_i(\boldsymbol Y_i,\boldsymbol X_i,A_i,\boldsymbol \beta, \boldsymbol \eta_W, \boldsymbol \eta_B) \\
 \boldsymbol S^W_i(\boldsymbol X_i, A_i, \boldsymbol \eta_W)\\
 \boldsymbol S^B_i(\boldsymbol X_i, A_i, \boldsymbol \eta_B)\\
 \end{array} \right),$$
where $\boldsymbol S^W_i$ and $\boldsymbol S^B_i$ represent the score equations for patients in cluster $i$ for the estimation of $\boldsymbol \eta_W$ and $\boldsymbol \eta_B$  in the PS and the OM. A standard Taylor expansion paired with Slutzky's theorem and the central limit theorem provide the sandwich estimator adjusted for nuisance parameters estimation in the OM and PS. We refer to this as the nuisance-adjusted sandwich estimator:
\begin{equation}
Var(\boldsymbol \Omega)={{E\left[\frac{\partial   \boldsymbol U_i(\boldsymbol \Omega)}{\partial \boldsymbol \Omega}\right]}^{-1}}^{T} \underbrace{{E\left[ \boldsymbol U_i(\boldsymbol \Omega)\boldsymbol U_i^T(\boldsymbol \Omega) \right]}}_{\boldsymbol \Delta_{adj}} \underbrace{E\left[\frac{\partial   \boldsymbol U_i(\boldsymbol \Omega)}{\partial \boldsymbol \Omega}\right]^{-1} }_{\boldsymbol \Gamma^{-1}_{adj}}.\label{eq:varnuis}
\end{equation} 
The variance estimator $\widehat{var}(\hat{\boldsymbol \beta}_{aug})$ is obtained by estimating unknown quantities upon substituting empirical means for expectations and $\widehat{\boldsymbol \Omega}=(\widehat{\boldsymbol \beta}, \widehat{\boldsymbol \eta_W}, \widehat{\boldsymbol \eta_B})$ for $\boldsymbol \Omega$. Thus, the term $\widehat{\Delta_{adj}}$ is given by  $ \frac{1}{M}\sum_{i=1}^M \widehat{\boldsymbol U_i}(\widehat{\Omega})\widehat{\boldsymbol U_i}(\widehat{\Omega})^T$ and $\widehat{\boldsymbol\Gamma_{adj}}$ is given by $ \frac{1}{M}\sum_{i=1}^M \frac{\partial   \widehat{\boldsymbol U_i}( \widehat{\boldsymbol \Omega})}{\partial  {\boldsymbol \Omega}}$.

In small sample settings, it is likely that this estimator of the variance of $\hat{\boldsymbol \beta}_{aug}$ is biased. We implemented Fay's bias-correction approach, which is particularly suitable for M-estimators (Fay et al. 2001). The term $\widehat{\boldsymbol \Delta_{adj}}$ in Equation \ref{eq:varnuis} is replaced by $\widehat{\boldsymbol \Delta_{fay}}$ given by ${\frac{1}{M}\sum_{i=1}^M\left[  \widehat{\boldsymbol H_i}\widehat{\boldsymbol U_i}(\widehat{\boldsymbol \Omega}) \left( \widehat{\boldsymbol H_i }\widehat{\boldsymbol U_i}(\widehat{\boldsymbol \Omega})\right)^T  \right]}$, where $ \widehat{\boldsymbol H_i}$ is a diagonal matrix with diagonal terms  $\widehat{{\boldsymbol H_i}}_{[jj]}=\left[ 1-min(q,(\frac{\partial  \widehat{ \boldsymbol U_i}(\widehat{\boldsymbol \Omega})}{\partial {\boldsymbol \Omega}} \widehat{\boldsymbol \Gamma^i_{adj}})_{[jj]} \right]$, $q=0.75$ is a frequently-used bound.

 \subsection{Definition of covariate interference and implication for analysis}
 \label{part:DAG}
 
 \label{part:supp2}
In previous sections, we discussed covariates measured on the index subject ($j$), but other subjects' ($j'$) covariates may also impact the outcome for the index subject. An example of a potentially interfering covariate is described by \citet{kaiser2011factors} who found a positive association between age of partner and infection with  HIV.   Similarly,  the characteristics of subgroups to which the index case belongs (household, neighborhoods, \dots), whether known or not, may be interfering covariates \citep{brumback2011adjusting}. In this paper, we consider the phenomenon of covariate interference where there exists at least one individual $j' \neq j$ such that   $E(Y_{ij}|\boldsymbol X_{ij}) \neq E(Y_{ij}|\boldsymbol X_{ij}, \boldsymbol X_{ij'})$, where $\boldsymbol X_{ij}$ represent the vector of all measured baseline covariates. That is, even after all covariates for the index subject $j$ have been included in the model, the covariates of individuals other than the index subject still affect the outcome of the index subject $j$; we refer to such covariates as interfering covariates.  See~\citet{sullivan1994cautionary} for a similar definition in longitudinal data and see~\citet{seaman2014review,liu2014large} for an analogous definition in non-randomized clustered data in the context of confounding by cluster and interference. Refer to Web-Supplementary Material Section A for a causal interpretation of covariate-interference. 

When interfering covariates affect either the outcome ($E(Y_{ij}|\boldsymbol X_{ij}) \neq E(Y_{ij}|\boldsymbol X_{ij},\boldsymbol X_{ij'})$) or the missingness process ($E(R_{ij}|\boldsymbol X_{ij}) \neq E(R_{ij}|\boldsymbol X_{ij}, \boldsymbol X_{ij'})$), but not both, the DR estimator is CAN even if the interfering covariates are not included in the models, provided that either the PS \textcolor{black}{($P(R_{ij} | \boldsymbol X_{ij},A_i)$)} or the OM \textcolor{black}{($E(Y_{ij} | \boldsymbol X_{ij},A_i=a)$)} is correctly specified; that is, either the PS or the OM includes all the covariates $\boldsymbol X_{ij}$ involved in the same functional form as in the data generation processes.  Accounting for covariates interference in the OM increases efficiency if and only if they predict the outcome.  When interfering covariates impact both the outcome and the missing data generating processes, they must be included in either the OM or the PS models in a way that correctly represents the data generation processes. Thus, it will ensure that the DR estimator will be CAN \textcolor{black}{if a correct model for either the OM ($E(Y_{ij} | \boldsymbol X_i,A_i=a)$) or the PS ($P(R_{ij} | \boldsymbol X_i,A_i)$) is specified, where the $\boldsymbol X_{ij}$ are replaced with $\boldsymbol X_i$ in the formulas above}. We acknowledge that this model for interfering covariates is not likely to be known and can be difficult to identify. Different cluster sizes and sub-clustering structures (such as households) may make infeasible the use of regression techniques in the OM or the PS  because of the potentially different dimensions of the individual and interfering covariates. Cluster summary measures such as the mean or maximum of individual covariates in the cluster (or sub-groups in each cluster) may nonetheless be useful in incorporating interference covariates in models \citep{brumback2010adjusting}. 

\section{Application}
\label{part:application}
\subsection{Description of the SAM study}
\label{part:descSAM}

We analyze data from the ``South African Men'' (SAM) study which randomized 22 pair-matched clusters  to a health-promotion intervention (control) and an HIV/STI risk-reduction intervention in a CRT design; the study included 1181 South African men who have sex with women. A complete description of the study design can be found in  \citep{jemmott2014cluster}. We focus on a cross-sectional analysis of these data after one year and ignore matching. The primary outcome of our analysis is  the overall percentage  of acts of protected intercourse among  the total number of acts of intercourse. When the total number of acts of intercourse is  zero, we set the percentage  to 100\%,  as no exposure implies no risk. Secondary outcomes are the percentages of protected acts of intercourse by type of partnership and type of intercourse (vaginal and anal sex with main and casual partners). Descriptive statistics for these outcomes, including proportion of missing observations by type of partner and intercourse  are provided in Table \ref{tab:descSAMoutcome}. Slightly more observations are missing in the HIV/STI intervention group (20.8\% versus 17.5\%). The overall protection percentage after one year are about 64\% for the HIV/STI intervention compared to 60\% for the control group. 

 As the proportion of missing baseline covariates was less than $0.1$\%, we consider them to be MCAR and exclude observation with missing covariates from the analysis. No community sub-structure, such as household or neighborhood structures, was described in the SAM study. Here we consider potential interfering covariates at a cluster level by taking the mean (or mode for qualitative variables) of  baseline measures in the community: $\overline{X^k_{i.}}=\frac{1}{n_i}\sum_{j=1,\dots,  n_i}  X^k_{ij}$. For example  \citet{hawkes2013hiv} demonstrated that the mean religiosity score for a community, defined as the mean of individual religiosity score in the community, may have an impact on each individual outcome and missingness in particular regarding sexual behaviors. Table \ref{tab:descSAMIndividual} describes socio-demographical individual-level variables and interfering covariates. We provide p-values for Wald tests testing the association of  covariates and treatment-covariate interactions with the outcome and the missingness indicator. In this study, there is evidence of interactions of individual covariates with treatment for both the outcome and the missing data generation processes. However, the interfering covariates defined here do not appear to be significantly associated  with both the outcome and the missing data generation process.

 \begin{table}[t]
\centering
\caption{Descriptive statistics of outcomes, sociodemographic individual covariates and interfering covariates by intervention group in SAM study. }
\small
\begin{tabular}{lcccc}
\hline 
 \multicolumn{5}{c}{\textbf{Descriptive Statistics of the outcomes} } \\
 \hline
\multicolumn{1}{c}{\textbf{}} &\multicolumn{2}{c}{HIV/STI}  &\multicolumn{2}{c}{Control group}  \\
\cline{2-5}
 \multicolumn{1}{c}{ }& Mean [IQR] & \% missing & Mean [IQR] & \% missing \\
 \hline
   \multicolumn{5}{l}{\textbf{Primary outcome for percentage of protection ($Y$)}} \\
   \hline
    Overall& 64\% [26; 100]& 20.8\% &60\% [22; 100] & 17.5\%\\
 \hline
    \multicolumn{5}{l}{\textbf{Secondary outcomes for percentage of protection ($Y^1, Y^2, Y^3$ and $Y^4$)}} \\
     \hline
     Main partner vaginal sex& 61\% [22; 100] &10.2\% &56\% [~~0; 100]&~ 9.3\%\\
 Casual partners vaginal sex& 68\% [33; 100] & 19.7\% & 68\% [33; 100] & 17.1\%\\
 Main partner anal sex& 37\% [~~0; ~68] & 11.2\% &52\% [~~0; 100]&~8.6\%\\
 Casual partners anal sex& 35\% [~~0; 100] & 15.1\% & 31\% [~~0; 100] & 12.8\%\\
 \hline 
\end{tabular}
\vspace{0.65cm}

\begin{tabular}{lcccccccc}
 \hline
  \multicolumn{9}{c}{\textbf{Descriptive Statistics of the covariates}}\\
 \hline
 \multicolumn{1}{c}{\textbf{}} &\multicolumn{2}{c}{\textbf{} } &&\multicolumn{5}{c}{\textbf{p-value for association with }} \\
\multicolumn{1}{c}{\textbf{}} &\multicolumn{1}{c}{HIV/STI}  &\multicolumn{1}{c}{Control group} &&\multicolumn{2}{c}{\textbf{Y}$^*$} &&\multicolumn{2}{c}{\textbf{P(Y observed)}$^{**}$} \\
\cline{2-5}\cline{5-6}\cline{8-9}
 \multicolumn{1}{c}{ }& Mean [IQR]  & Mean [IQR]  & & $\eta^O_2\neq0$ & $\eta^O_3\neq0$ & & $\eta^M_2\neq0$ & $\eta^M_3\neq0$ \\
 \hline
  \multicolumn{9}{l}{\textbf{Individual covariates $X_{ij}$}} \\
    \hline
    Age & 26 [21; 30]  & 26.5 [21; 31] && 0.41 & {0.13} && \textbf{0.03} & 0.18 \\ 
 Employment Yes & 23\%  & 26\% && \textbf{0.04} & 0.17 && \textbf{0.01} & \textbf{$<$0.001} \\ 
 Married Yes & 23\%  & 24\% && \textbf{0.05} & 0.76 && 0.68 & 0.50 \\ 
 Education Yes & 46\%  & 42\%  && 0.58 & \textbf{$<$0.001} && 0.76 & \textbf{0.05} \\ 
 Number of children & 1.5 [0; 2]  & 1.7 [0 ;2]  && 0.21 & {0.12} && 0.25 & 0.31 \\ 
 Wealth & 5.3 [4; 7]  & 5.3 [4; 7]  && 0.77 & 0.96 && 0.25 & 0.54 \\ 
 Social desirability & 3.4 [3.2; 3.4]  & 3.4 [3.2; 3.4]  && 0.87 & 0.33 && \textbf{0.04} & 0.34 \\ 
 Religiosity & 0.01 [-0.7 ;0.7]  & 0.00[-0.7 ;0.6]  && 0.46 & 0.25 && \textbf{0.07} & 0.69 \\ 
 HIV/STI Knowledge & 14.3 [12; 17]  & 14.1 [12; 17]  && {0.13} & 0.93 && 0.37 & \textbf{0.03} \\ 
  Condom Behaviors & 3.7 [3.3 ;4]  & 3.7 [3.3 ;4.1]  && \textbf{$<$0.001} & 0.36 && 0.16 & 0.33 \\ 
 Condom Knowledge & 3.1 [3; 4]  & 3.1 [3; 4]  && 0.41 & 0.57 && 0.21 & \textbf{0.06} \\ 
 Condom Efficacy & 3.9 [3.7 ;4.2]  & 3.9 [3.7 ;4.2]  && \textbf{0.01} & 0.31 && 0.97 & 0.42 \\ 
 Condom Peer norm & 3.7 [3.4 ;4.1]  & 3.7 [3.4 ;4]  && \textbf{$<$0.001} & 0.71 && 0.49 & 0.32 \\ 
 Never had HIV test & 20\%  & 21\%  && 0.61 & 0.80 && 0.74 & 0.34 \\ 
 Sexual Activity Yes & 84\%  & 84\%  && 0.71 & \textbf{0.06} && 0.53 & 0.77 \\ 
 Eating attitude & 4.2 [4 ;5]  & 4.2 [3.7 ;5]  && 0.76 & \textbf{0.01} && 0.74 & 0.53 \\ 
 Exercise Yes & 43\%  & 42\%  && 0.99 & \textbf{0.04} && {0.12} & 0.46 \\ 
 CAGE $>=$ 2 & 62\%  & 58\% & & 0.22 & 0.41 && 0.18 & \textbf{0.08} \\ 
 Health Knowledge & 10.8 [9; 12]  & 10.6 [9; 13]  && 0.51 & 0.38 && 0.59 & 0.83 \\ 
 \hline 
  \multicolumn{9}{l}{\textbf{Interfering covariates $\overline{X_{i.}}=\frac{1}{n_i}\sum_{j=1,\dots,  n_i} X_{ij}$}} \\
   \hline
     Mean Age & 26 [25 ;27]  & 27 [26 ;28]  && 0.39 & 0.96 && \textbf{0.05} & \textbf{0.10} \\ 
 Mean Education Yes & 27\%  & ~~8\%  && 0.58 & 0.61 && 0.72 & 1.00 \\ 
 Mean Number of children & 1.6 [1.2; 2.1]  & 1.7 [1.1 ;2.1]  && 0.81 & 0.67 && {0.14 }& 0.59 \\ 
 Mean Wealth & 5.4 [4.4 ;6.2]  & 5.2 [4.4 ;6.1]  && 0.45 & 0.38 && 0.23 & 0.92 \\ 
 Mean  Social desirability  & 3.4 [3.3 ;3.4]  & 3.4 [3.3 ;3.4]  && 0.16 & 0.44 && 0.60 & 0.85 \\ 
 Mean Religiosity & 0.00 [-0.1 ;0.1]  & 0.00 [-0.1 ;0.1]  && 0.84 & 0.70 && 0.18 & 0.94 \\ 
 Mean HIV/STD Knowledge & 14.2 [14; 15]  & 13.9 [13 ;14]  && 0.37 & 0.23 && \textbf{0.01} & 0.45 \\ 
 Mean Condom Behaviors & 3.7 [3.6 ;3.8]  & 3.7 [3.7 ;3.8] & & 0.37 & 0.40 && \textbf{0.02} & 0.95 \\ 
 Mean Condom Knowledge & 3.1 [2.9 ;3.3]  & 3.1 [2.9 ;3.2]  && 0.52 & 0.21 && {0.15} & 0.32 \\ 
 Mean Condom Efficacy & 3.9 [3.7 ;4.0]  & 3.9 [3.8 ;4.0]  && 0.23 & 0.38 && 0.21 & 0.58 \\ 
 Mean Condom peer norm & 3.7 [3.6 ;3.8]  & 3.7 [3.6 ;3.7]  && 0.23 & 0.52 && \textbf{$<$0.001} & \textbf{0.01} \\ 
 Mean Eating attitude & 4.2 [4.1;4.3]  & 4.2 [4.0 ;4.3]  && 0.71 & {0.15} && 0.25 & \textbf{0.07} \\ 
 Mean Exercise Yes & 76\%  & 82\% & & 0.43 & 0.53 && \textbf{0.10} & 0.82 \\ 
 Mean CAGE$>$=2 & 63\%  & 37\%  && 0.99 & 0.79 && 0.71 & 0.41 \\ 
 Mean Health Knowledge & 10.7 [10.5 ;11]  & 10.6 [10.3 ;10.8]  && \textbf{0.10} & \textbf{0.10} && {0.15} & 0.73 \\ 
 \hline
 \multicolumn{9}{l}{\textbf{$^*$} Wald test for $\eta^O_2$ and $\eta^O_3$ in the regression $Y=\eta^O_0+ \eta^O_1 A +\eta^O_2 X +\eta^O_3 AX$}\\
 \multicolumn{9}{l}{\textbf{$^{**}$} Wald test for $\eta^M_2$ and $\eta^M_3$ in the regression $logit[P(R=1)]=\eta^M_0+ \eta^M_1 A +\eta^M_2 X +\eta^M_3 AX$}\\
 \hline 
\end{tabular}

\label{tab:descSAMIndividual}
\label{tab:descSAMinterf}
\label{tab:descSAMoutcome}
\end{table}

\subsection{Results} 
\label{part:tabSAM} 

We analyze these data with the GEE, the AUG, the IPW and the DR using both independence (-I) and exchangeable (-E) working correlation structures.
Variables for the PS, and the OM were selected using a forward stepwise regression (separately for each  treatment group) from among all the individual covariates $\boldsymbol X_{ij}$ presented in Table \ref{tab:descSAMinterf}. We did not include the interfering covariates ($\overline{X_{i.}}$) in the analysis as none impacted both  outcome and  missingness processes (Table \ref{tab:descSAMinterf}). We used the \textit{step} function in R based on the AIC criterion. Results of these selections are given in Web-Supplementary Material F. We describe here the results for the primary outcome. The amount of missingness is larger  in the treated arm and increases with age; it decreases with religiosity, good health score, and exercise. The OM patterns are substantially different for treated and control; the only common variable  is the CAGE score.  In both arms lower alcohol consumption is associated with  a greater percentage of protected acts of intercourse.  Results are presented in Table \ref{tab:resultSAM1} for primary and secondary outcomes. With the DR-E, we observe a significant difference of 7.4\% (sd=2.9\%, p=0.01) in the overall percentage of protected intercourse in the HIV/STI intervention group compared to the control group. Analyses of the secondary outcomes suggest that this result is mainly driven by condom use during vaginal intercourse with a marital partner. The HIV/STI intervention has no significant impact on other outcomes. 
Using the DR rather than the standard GEE or the AUG has an impact on the treatment effect estimates and associated standard errors (SE). The difference between these approaches is  apparent in the magnitude and direction of the marginal treatment effect estimate. For example, the analysis for the GEE-I (3.8 [-1.0; 8.5]) does not demonstrate a significant effect of  the HIV/STI intervention on   overall percentage of protected intercourse, whereas this effect is stronger and significant for the DR-I (7.3 [1.6; 13.0]). Both the GEE-I and the AUG-I (5.4 [2.2; 8.7]) are probably biased due to missing data. Using the DR instead of the IPW leads to an increased magnitude of the treatment effect and an increased level of statistical significance: for example, the DR-E (7.4 [1.73; 13.0]) compared to the IPW-E (3.4 [-1.4; 8.3]).

\begin{table}[ht]
\centering
\caption{Analysis of effect of STI/HIV intervention on overall percentage of protected intercourses during the last 3 months one year after intervention (primary outcome) and stratified by intercourse types (secondary outcomes)  in SAM study with the GEE, the IPW, the AUG and the DR. }
\begin{tabular}{lccccccc}
 \hline
 &\multicolumn{3}{c}{\textbf{Independence (-I)}}&~~~~~ &\multicolumn{3}{c}{\textbf{Exchangeable (-E)}}  \\
 &$\hat{\beta}_A$& SE & p-value~~ && ~~$\hat{\beta}_A$ & SE & p-value\\
\hline
 \multicolumn{8}{l}{\textbf{Overall percentage of protected intercourse ($Y$)}}\\
 \hline
   GEE  & 3.751 & 2.419 & 0.121 && 3.738 & 2.361 & 0.113 \\ 
  IPW  & 3.445 & 2.558 & 0.178 && 3.429 & 2.488 & 0.168 \\ 
  AUG  & 5.414 & 1.665 & 0.001 && 5.478 & 1.633 & 0.001 \\ 
   DR~~~~~~~~~~~  & 7.341 & 2.923 & 0.012 && 7.386 & 2.885 & 0.010 \\ 
 \hline
 \multicolumn{8}{l}{\textbf{Percentage of protected vaginal intercourse with marital partner ($Y^1$)}}\\
 \hline
   GEE  & 5.805 & 2.689 & 0.031 && 5.761 & 2.67 & 0.031 \\ 
  IPW  & 5.660 & 2.720 & 0.037 && 5.626 & 2.698 & 0.037 \\ 
  AUG  & 6.550& 1.811 & $<$0.001 && 6.518 & 1.794 & $<$0.001 \\ 
  DR  & 7.254 & 2.542 & 0.004 && 7.273 & 2.50 & 0.004 \\ 

 \hline
 \multicolumn{8}{l}{\textbf{Percentage of protected vaginal intercourse with casual partner ($Y^2$)}}\\
 \hline
   GEE  & -0.621 & 4.180 & 0.882 && -0.497 & 4.164 & 0.905 \\ 
  IPW  & -1.500 & 4.182 & 0.720 && -1.356 & 4.17 & 0.745 \\ 
  AUG  & -1.191 & 2.638 & 0.652 && -1.121 & 2.624 & 0.669 \\ 
  DR  & -2.103 & 4.077 & 0.606 && -2.018 & 4.058 & 0.619 \\ 
 \hline
 \multicolumn{8}{l}{\textbf{Percentage of protected anal intercourse with marital partner ($Y^3$)}}\\
 \hline
 GEE  & -0.983 & 1.083 & 0.364 && -0.972 & 1.081 & 0.369 \\ 
  IPW  & -0.934 & 1.087 & 0.390 && -0.921 & 1.085 & 0.396 \\ 
  AUG  & -0.951 & 0.684 & 0.164 && -0.954 & 0.684 & 0.163 \\ 
  DR  & -0.835 & 1.005 & 0.406 && -0.819 & 1.003 & 0.414 \\ 
 \hline
 \multicolumn{8}{l}{\textbf{Percentage of protected anal intercourse with casual partner ($Y^4$)}}\\
 \hline
   GEE  & 0.013 & 1.201 & 0.991 && -0.002 & 1.204 & 0.998 \\ 
  IPW  & -0.003 & 1.181 & 0.998 && -0.019 & 1.184 & 0.987 \\ 
  AUG  & -0.467 & 0.834 & 0.576 && -0.476 & 0.837 & 0.570 \\ 
  DR  & -0.963 & 1.207 & 0.425 && -0.971 & 1.208 & 0.421 \\ 
 \hline
\end{tabular}
\label{tab:resultSAM2}
\label{tab:resultSAM1}
\end{table}

\section{Simulation Studies}
\label{part:simul}
\subsection{Properties of the DR estimator}
\label{part:toyinterfX1}
We consider a setting with continuous outcome $\boldsymbol Y_{ij}$ and assignment of treatment $A_i$ at a cluster level with probability $p=1/2$. We generate a normally distributed covariate $X1_{ij}$ (independent of $A_i$) with mean 1 and a standard deviation of 5. For each individual, we define a covariate $\overline{\boldsymbol X1_{i.}}$ which is the mean of $\boldsymbol X1$ for all the subjects in the same cluster: $\overline{\boldsymbol X1_{i.}}=\frac{1}{n_i}\sum_{j=1}^{n_i}  X1_{ij}$. Similarly, we generate $ X2_{ij}\sim\mathcal{N}(2,5)$ and $ X3_{ij} \sim\mathcal{N}(3,5)$; $\overline{\boldsymbol X2_{i.}}$ and $\overline{\boldsymbol X3_{i.}}$ are defined as was $\overline{\boldsymbol X1_{i.}}$ and are possible interfering covariates. The model for simulation is given in Equation \ref{eq:simulgeneration}:

\vspace{-0.4cm}
 \begin{eqnarray}
 \hspace{-0.5cm}
\left\lbrace \begin{array}{ccl}
Y_{ij} &= & \beta^O_{0} + \beta^O_{A} A_i + \beta^O_{1} X1_{ij}+  \beta^O_{I1}  \overline{X1}_{i.} +  \beta^O_{A1} A_i X1_{ij}+\epsilon^O_{i}+\epsilon^O_{ij} \\
logit(P(R_{ij}=0)) &=& \beta^M_{0}  +  \beta^M_{A}  A_i +  \beta^M_{1} X1_{ij} + \beta^M_{I1}  \overline{X1}_{i.}+  \beta^M_{A1}  A_i X1_{ij}
\end{array} \right. . \label{eq:simulgeneration}
 \end{eqnarray}
 \normalsize
 
\vspace{0.2cm}
 The parameters $\boldsymbol \beta^O=(\beta^O_{0}, \beta^O_{A},  \beta^O_{1},  \beta^O_{I1},  \beta^O_{A1})$ are the regressors associated with intercept, treatment, covariate, interfering covariate, treatment-covariate interaction for the outcome model. Parameters $\boldsymbol \beta^M$ are the same for the missing data generating process. Scenarios with low correlation among cluster ($0.05$) were simulated with $\epsilon^O_{i}\sim\mathcal{N}(0,0.05)$ and $\epsilon^O_{ij}\sim\mathcal{N}(0,1.0)$ for cluster and individual random errors; scenarios with high correlation ($0.2$)  were simulated with $\epsilon^O_{i}\sim\mathcal{N}(0,0.25)$ and $\epsilon^O_{ij}\sim\mathcal{N}(0,1.0)$. True correlation structure is exchangeable. We investigate small sample ($M=10$ and $n_i=(10,20,30)$ with probability 1/3 each) and large sample ($M=100$ and $n_i=(90,100,110)$ with probability 1/3 each) properties. In each scenario, we generate 1000 replicates of datasets.

 We evaluate the double robustness of the DR estimator in the setting of large and small sample with low correlation, but similar results are observed for large correlation.  We investigate models of analysis with OM and PS correctly specified (TRUE), misspecified (MISS) and partially specified omitting treatment-covariate interactions (NONE). Table \ref{tab:bothDR}describes the data generation process, provides the formulations of the models of analysis, and shows the results from analysis; on average, 26\% of outcomes were missing and the  average ICC was 0.08. When there is no missing data, the traditional GEE is consistent because of randomization. When outcome data are MAR but not MCAR, the GEE and the AUG analysis are biased (-1.7 for the GEE-I and -1.8 for the AUG-I). When either the OM  or the PS  models or both are correctly specified there is negligible estimated bias for the DR - a finding that confirm consistency. In small samples, this bias is bigger when only the PS is correct because the weights are estimated with lower accuracy. Using the more common choice of  implementation for the weighted GEE $\boldsymbol W_i^{1/2}(\boldsymbol \eta_W) \boldsymbol  V_i^{-1} \boldsymbol  W_i^{1/2}(\boldsymbol \eta_W)$ leads to very high bias if an exchangeable correlation structure is used (0.374 if the OM is correct and 858 if it is not, for large sample). When the OM is correct the coverage remains around 95\% (see Table 2 in Web-Supplementary Material E). Using $\boldsymbol  V_i^{-1} \boldsymbol  W_i(\boldsymbol X_i, A_i, \boldsymbol \eta_W)$ in the implementation of weights addresses this problem and permits the use of correlation structures other than independence. The IPW with correct PS also corrects the bias (-0.01) but is less efficient than the DR approach; coverage is close to the nominal value of 95\%. In small samples, the empirical SE are underestimated. By contrast, in the large sample setting, using the nuisance-adjusted sandwich estimator for the DR leads to good estimates of the asymptotic SE (0.0263) compared to the empirical SE (0.0266) over 1000 replicates. Moreover, we observe that the coverage using the DR is comparable to that of the GEE with complete data. Finally, we note that when the treatment-covariate interactions are ignored in the PS and only accounted for in the OM by fitting a different regression in each treatment group, the DR approach is also consistent and achieve same precision as when both the PS and the OM are correct (0.0014 and SE=0.027 for OM.TRUE.PS.NONE and 0.0013 SE=0.029 for OM.TRUE.PS.TRUE).

\begin{table}[ht]
\centering
\caption{Properties for the Doubly robust estimator (DR) compared to the GEE, the IPW and the AUG using the data generation mechanism from Equation \ref{eq:simulgeneration} with covariate interference for the outcome and missing data generation process. Misspecified (.MISS), correctly specified (.TRUE) and partially specified without treatment-covariate interactions (.NONE) OM and PS are investigated. Statistics for 1000 replicates are the bias compared to $M_E^*=2.0$, the empirical standard errors over the replicates, the mean asymptotic nuisance-adjusted standard error} and the coverage with independence (-I) and exchangeable (-E) working correlation matrix. 
\small
\begin{tabular}{lccccccccc}
 \hline
  &&\multicolumn{2}{c}{\textbf{ }} &\multicolumn{4}{c}{\textbf{Standard Error (SE)}} &\multicolumn{2}{c}{\textbf{Coverage}} \\
 &&\multicolumn{2}{c}{\textbf{Bias}} &\multicolumn{2}{c}{\textbf{Empirical }} &\multicolumn{2}{c}{\textbf{Robust}}  &\multicolumn{2}{c}{\textbf{95\%}} \\
 &$\boldsymbol{M_E^*}$ & -I & -E & -I & -E & -I & -E & -I & -E  \\
            \hline
            \multicolumn{10}{l}{\textbf{Small sample $M=10, n_i=(10,20,30)$} with probability 1/3 each, \textbf{Low correlation}}\\
 GEE (no missing) & 2.0& 0.0186 & 0.0171 & 0.6553 & 0.6598 &0.5629 & 0.5682 & 93.0 & 92.9  \\ 
 \hline
  GEE  &2.0 & -1.7186 & -1.7166 & 0.5717 & 0.5724 &   0.5074 & 0.4306  & 12.8  & 7.2  \\ 
  IPW.PS.TRUE  &2.0&  -0.1623 & -0.1689 & 1.1447 & 1.1473 & 0.7987 & 0.8161 &  83.9 & 84.7 \\ 
    AUG.OM.TRUE  &2.0& -1.8142 & -1.8134 & 0.4530 & 0.4148 & 0.8751 & 0.8699 & 39.4 & 38.0 \\ 
 \hline
     DR.OM.MISS.PS.TRUE &2.0& -0.0127 & -0.0366 & 2.7327 & 2.6793 & 1.4029 & 1.3985 &  92.0 & 92.0 \\ 
     DR.OM.TRUE.PS.MISS &2.0& 0.0011 & 0.0001 & 0.1544 & 0.1545 & 0.1287 & 0.1330 &  86.0 & 87.5  \\
     DR.OM.TRUE.PS.TRUE &2.0& -0.0017 & -0.0022 & 0.1881 & 0.1838 & 0.1413 & 0.1447 &  86.9 & 87.4   \\
    \hline 
   DR.OM.TRUE.PS.NONE&2.0& 0.0006 & -0.0003 & 0.1612 & 0.1608 & 0.1330 & 0.1368 & 85.8 & 87.8  \\
\hline
\hline
            \multicolumn{10}{l}{\textbf{Large sample $M=100, n_i=(90,100,110)$} with probability 1/3 each, \textbf{Low correlation}}\\
 GEE (no missing) & 2.0& 0.0042 & 0.0043 & 0.1156 & 0.1157 & 0.1155  & 0.1156  & 94.3  & 94.5 \\
 \hline
  GEE  &2.0 & -1.7335 & -1.7321 & 0.1015 & 0.1013 &0.0994  &0.0994   &  0.0 & 0.0  \\
  IPW.TRUE  &2.0&  -0.0113 & -0.0108 & 0.2626 & 0.2621 & 0.2507 & 0.2510  & 93.5 & 93.9   \\
  AUG.TRUE  &2.0& -1.8021 & -1.8024 & 0.0694 & 0.0664 & 0.2556 & 0.2550 & 0.0 & 0.0 \\
 \hline
    OM.MISS.PS.TRUE &2.0& -0.0089 & -0.0079 & 0.3127 & 0.3105 & 0.3937 & 0.3940 & 99.3 & 99.1 \\
    OM.TRUE.PS.MISS  &2.0& 0.0013 & 0.0014 & 0.0259 & 0.0259 & 0.0256 & 0.0257  & 95.2 & 95.7  \\
  OM.TRUE.PS.TRUE&2.0& 0.0013 & 0.0014 & 0.0284 & 0.0284 & 0.0285 & 0.0285  & 95.8 & 96.0 \\
    \hline 
  OM.TRUE.PS.NONE&2.0&  0.0014 & 0.0014 & 0.0266 & 0.0266 & 0.0263 & 0.0263  & 95.2 & 95.1 \\
  \hline
    \multicolumn{10}{l}{\textbf{Marginal model for the GEE:}}\\
  \multicolumn{10}{l}{$ \qquad \qquad  \mu(\boldsymbol \beta, A_i)= \beta_0+ \beta_A  A_i$}\\
 \multicolumn{10}{l}{\textbf{OM is fitted for each treatment group $A_i=a$:}}\\
  \multicolumn{10}{l}{$  \qquad \qquad \text{OM.TRUE~~~~~}  B(\boldsymbol X_i, A_i=a)= \gamma^a_0 +  \gamma^a_1  X1_{ij}+ \gamma^a_2  \overline{ X1}_{i.}  $}\\
   \multicolumn{10}{l}{$\qquad \qquad \text{OM.MISS~~~~~~}  B(\boldsymbol X_i, A_i=a)= \gamma^a_0 +  \gamma^a_1  X2_{ij} $}\\
  \multicolumn{10}{l}{\textbf{PS is fitted for the whole dataset:}}\\
  \multicolumn{10}{l}{$  \qquad \qquad \text{PS.TRUE~~~~~~}  \pi_{ij}(\boldsymbol X_i, A_i)= expit \left( \gamma^M_0 +  \gamma^M_A  A_{i} +  \gamma^M_1  X1_{ij}+ \gamma^M_2  \overline{ X1}_{i.} + \gamma^M_3  A_{i}   X1_{ij} \right)  $}\\
   \multicolumn{10}{l}{$\qquad \qquad \text{PS.MISS~~~~~~~}  \pi_{ij}(\boldsymbol X_i, A_i)= expit \left( \gamma^M_0 +  \gamma^M_A  A_{i} +  \gamma^M_1  X2_{ij}\right) $}\\
    \multicolumn{10}{l}{$\qquad \qquad \text{PS.NONE~~~~~}  \pi_{ij}(\boldsymbol X_i, A_i)= expit \left( \gamma^M_0 +  \gamma^M_A  A_{i} +  \gamma^M_1  X1_{ij}+ \gamma^M_2  \overline{ X1}_{i.}\right)$}\\
\hline
\end{tabular}
\label{tab:bothDR}
\end{table}

Table \ref{tab:both} presents the results of analyses with the GEE, the IPW, the AUG and the DR that investigate the impact of correlation of the outcome in the data with small and large sample. 
The average percentage of missing outcomes is 23\%; the average ICC is 0.04 for low correlation and 0.21 for high correlation. We analyzed the data using a PS and an OM model  that was fit using a stepwise variable selection from among all of the individual and interfering covariates described above. The GEE and the AUG estimates are systematically biased because there is no correction for missing data. The IPW is also biased because the PS is incorrect in that  it omits treatment-covariate interactions. The DR estimates are  consistent in all analyses. In small sample settings, the empirical SE is underestimated even when using nuisance-adjusted SE, but estimation  is  improved by Fay's correction. Nonetheless,  the coverage remained lower than 86\%, but it improves for large samples. Finally, when there is low correlation in the outcome, the robust SE better approximate the empirical SE. 

\begin{table}[ht]
\centering
\caption{Sample size effect and correlation magnitude effects for data generation mechanism given in Equation \ref{eq:simulgeneration} with $\boldsymbol \beta^O=(1,1,1,1,1) $ and $\boldsymbol \beta^M=(-3,1/2,1/2,1/2,1/2)$. Statistics for 1000 replicates are the bias compared to $M_E^*$, the empirical standard errors over the replicates, the mean asymptotic nuisance-adjusted standard errors and the coverage for the GEE, the IPW, the AUG and the DR with independence (-I) and exchangeable (-E) working correlation matrix. }
\small
\begin{tabular}{lccccccccccccc}
  \hline
  &&\multicolumn{2}{c}{\textbf{ }} &\multicolumn{6}{c}{\textbf{Standard Error (SE)}} &\multicolumn{4}{c}{\textbf{Coverage}} \\
 &&\multicolumn{2}{c}{\textbf{Bias}} &\multicolumn{2}{c}{\textbf{Empirical }} &\multicolumn{2}{c}{\textbf{Robust}}  &\multicolumn{2}{c}{\textbf{Fay's}} &\multicolumn{2}{c}{\textbf{Robust}} &\multicolumn{2}{c}{\textbf{Fay's}}\\
 &$\boldsymbol{M_E^*}$ & -I & -E & -I & -E & -I & -E & -I & -E & -I & -E & -I & -E  \\
  \hline
            \multicolumn{14}{l}{\textbf{Small sample $M=10, n_i=(10,20,30)$} with probability 1/3 each, \textbf{Low correlation}}\\
  GEE~~~~~~~~~~~ &2.0 & -1.7473 & -1.7479 & 0.4351 & 0.4360  & 0.3963 & 0.3256 & 0.4559 & 0.4603   & 0.8 & 2.3  & 3.9 & 4.9  \\ 
  IPW &2.0& -1.0130 & -1.0130  & 0.6793 & 0.6842  & 0.5538 & 0.5591  & 0.6735 & 0.6766  &  49.0 & 49.2  & 59.8 & 59.9   \\
  AUG &2.0& -1.8099 & -1.8111  & 0.3371 & 0.3269  & 0.8362 & 0.8353  & 0.8834 & 0.8817  &  29.7 & 29.1 & 40.1 & 39.2   \\
  DR &2.0 & 0.0008 & 0.0006  & 0.1552 & 0.1586  & 0.1127 & 0.1140  & 0.1190 & 0.1201  & 84.8 & 83.8  & 86.0 & 86.2  \\
    \hline
           \multicolumn{14}{l}{\textbf{ Large sample $M=100, n_i=(90,100,110)$} with probability 1/3 each, \textbf{Low correlation}}\\
 GEE & 2.0&-1.7335 & -1.7321  & 0.1015 & 0.1013  & 0.0985 & 0.0727  & 0.0994 & 0.0994  & 0.0 &  0.0 & 0.0 & 0.0  \\
  IPW & 2.0&-0.9955 & -0.9952  & 0.1514 & 0.1517  & 0.1559 & 0.1563  & 0.1588 & 0.1592  & 0.2 & 0.2  & 0.2 & 0.2   \\
  AUG & 2.0&-1.8019 & -1.8022  & 0.0695 & 0.0664  & 0.2556 & 0.2550  & 0.2569 & 0.2563  & 0.0 & 0.0 & 0.0 & 0.0  \\
  DR & 2.0&0.0016 & 0.0017  & 0.0265 & 0.0265  & 0.0262 & 0.0263  & 0.0264 & 0.0264   & 95.1 & 95.0  & 95.1 & 95.2  \\
 \hline
             \multicolumn{14}{l}{\textbf{Small sample $M=10, n_i=(10,20,30)$} with probability 1/3 each, \textbf{High correlation}}\\
  GEE~~~~~~~~~~~ &2.0 &-0.0086 & -0.0086  & 0.5265 & 0.5314  & 0.4701 & 0.4721  & 0.5651 & 0.5657  & 88.5 & 88.4 & 92.9 & 92.7  \\
  IPW &2.0&  -1.0221 & -1.0229  & 0.7026 & 0.7083   & 0.5776 & 0.5829  & 0.7015 & 0.7044  & 52.4 & 52.2  & 62.2 & 61.5  \\
  AUG &2.0&-1.7985 & -1.7987  & 0.5058 & 0.5084  & 0.8748 & 0.8727  & 0.9243 & 0.9209   & 35.8 & 35.8  & 45.1 & 45.5  \\
  DR &2.0& 0.0098 & 0.0062  & 0.4328 & 0.4407  & 0.2469 & 0.2480  & 0.2607 & 0.2614   & 77.4 & 77.7  & 79.7 & 79.6   \\
    \hline
           \multicolumn{14}{l}{\textbf{Large sample $M=100, n_i=(90,100,110)$} with probability 1/3 each, \textbf{High correlation}}\\
GEE &2.0&  -1.7325 & -1.7312  & 0.1145 & 0.1141  & 0.1121 & 0.0753  & 0.1132 & 0.1132  & 0.0 & 0.0 & 0.0 & 0.0  \\
  IPW&2.0 &-0.9945 & -0.9940  & 0.1618 & 0.1620  & 0.1652 & 0.1656   & 0.1682 & 0.1686  & 0.2 & 0.2  & 0.2 & 0.2 \\
    AUG &2.0& -1.8014 & -1.8017  & 0.0787 & 0.0761   & 0.2587 & 0.2581 & 0.2600 & 0.2594  & 0.0 & 0.0 & 0.0 & 0.0  \\
   DR &2.0& 0.0029 & 0.0032  & 0.0609 & 0.0610   & 0.0590 & 0.0590  & 0.0593 & 0.0593  & 94.7 & 94.6  & 94.7 & 94.6  \\
  \hline
    \multicolumn{14}{l}{\textbf{Marginal model for the GEE:}}\\
  \multicolumn{14}{l}{$ \qquad \qquad  \mu(\boldsymbol \beta, A_i)= \beta_0+ \beta_A A_i$}\\
 \multicolumn{14}{l}{\textbf{OM in AUG and DR is fitted for each treatment group $A_i=a$ using a stepwise regression:}}\\
     \multicolumn{14}{l}{$\qquad \qquad  B(\boldsymbol X_i, A_i=a)= \text{stepwise}( X1_{ij}, X2_{ij}, X3_{ij},\overline{X1}_{i.},\overline{X2}_{i.},\overline{X3}_{i.}) $}\\
  \multicolumn{14}{l}{\textbf{PS in DR and IPW is fitted for the whole dataset using a stepwise regression:}}\\
     \multicolumn{14}{l}{$\qquad \qquad   logit(\pi_{ij}(\boldsymbol X_i, A_i))= \text{stepwise}(\boldsymbol A_i,  X1_{ij}, X2_{ij}, X3_{ij},\overline{X1}_{i.},\overline{X2}_{i.},\overline{X3}_{i.}) $}\\
\hline
\end{tabular}
\label{tab:both}
\end{table}

\subsection{Simulations mimicking the SAM Study}
\label{part:mimick}
To consider more complex settings, we mimic the SAM study (see Section \ref{part:application}). We simulate the following individual-level covariates: employment (${\rm EMP}\sim \mathcal{B}(0.25)$), marital status (${\rm MARI} \sim \mathcal{B}(0.23)$), age (${\rm AGE}\sim\mathcal{N}(27; 7)$), religiosity (${\rm REL}\sim\mathcal{N}(0, 0.8)$), the CAGE score (from a multinomial of probabilities ${\rm CAGE} \sim\mathcal{M}(0.3; 0.1; 0.1; 0.2; 0.3)$ for modalities 0,1,2,3 and 4), the HIV score (${\rm HIV} \sim \mathcal{N}(14; 4)$) and the condom knowledge score (${\rm CDM}\sim\mathcal{N}(3; 1)$). Interfering covariates are generated as means for quantitative variables or modes for qualitative variables of the individual-level variables in each of the community (as was done for $\overline{\boldsymbol X1}$, $\overline{\boldsymbol  X2}$ and $\overline{\boldsymbol  X3}$ in Section \ref{part:toyinterfX1}). We generate data from the model in  Equation \ref{eq:realsimul}. In simulating the outcome, we add cluster random errors to create an exchangeable correlation structure with $\epsilon^O_{i}\sim\mathcal{N}(0,5)$ and an individual random effects  $\epsilon^O_{ij}\sim\mathcal{N}(0,4)$. This provides an outcome correlation among clusters of $0.07$. We analyzed the data using a PS and an OM composed of all the covariates described above with a stepwise variable selection. Table \ref{tab:simulreel} shows the bias,  SE, and  coverage of the methods we consider based on  1000 replicates for the estimation of the  parameter $M_E^*=5.73$. The percentage of missing outcomes is 21\%  and the average empirical ICC is 0.06. 
\begin{eqnarray} 
\scriptstyle
 \left\lbrace \begin{array}{ccl} 
   \scriptstyle Y_{ij} &= & \scriptstyle  60 + 40A_i -9.0 {\rm EMP}_{ij} -8.0 {\rm MARI}_{ij}+ 1.0 {\rm CDM}_{ij} + 5.0 {\rm REL}_{ij} \\
& & + \underbrace{ \scriptstyle  A_i \left[ -2.0 {\rm AGE}_{ij} + 8.5 {\rm EMP}_{ij} + 3.5 {\rm MARI}_{ij}  + 1.5 {\rm HIV}_{ij} -2.0 {\rm CAGE}_{ij} + 2.0 {\rm REL}_{ij} \right]}_{\text{Interactions}}\\
& & \underbrace{ \scriptstyle  -0.5 \overline{{\rm AGE}_{i.}}- 7.0 \overline{{\rm CDM}_{i.}}- 5 \overline{{\rm REL}_{i.}}+ 1.0 \overline{{\rm HIV}_{i.}} }_{\text{covariate interference}} \scriptstyle  +\epsilon^O_{i}+\epsilon^O_{ij} \\
\scriptstyle  logit[P(R_{ij}=0)] &= & \scriptstyle  -3.0 + 2.0 A_i + 0.01 {\rm AGE}_{ij} - 0.1 {\rm HIV}_{ij} + \underbrace{ \scriptstyle  A_i \left[ -0.1 {\rm AGE}_{ij} -0.2{\rm HIV}_{ij} \right]}_{\text{Interactions}}\\
& & + \underbrace{\scriptstyle   0.02 \overline{{\rm AGE}_{i.}}+ 0.2\overline{{\rm CDM}_{i.}}+ 0.2 \overline{{\rm CAGE}_{i.}}}_{\text{covariate interference}} \scriptstyle  \\
 \end{array} \right. \label{eq:realsimul} 
   \end{eqnarray}
   \normalsize 

 Table \ref{tab:simulreel} provides the estimates the marginal treatment effect for small sample and for the same sample size as that of the SAM data. The GEE, the AUG and the IPW  yield biased results whereas the DR has small bias justifying its use to analyse the data even ignoring covariate interference. Fay's correction with coverage around 92\% in small sample and 95\% in large sample achieve good accuracy. Figure 2 in Web-Supplementary Material C3 represents the histograms of estimates over the 1000 replicates together with the true value of marginal treatment effect. It displays the bias of the GEE, the AUG and the IPW estimators compared to the DR and supports the approximate normal distribution of the DR estimator. 

\begin{table}[ht]
\centering
\caption{Simulation of the scenario described in Equation \ref{eq:realsimul} mimicking the SAM study data. Statistics for 1000 replicates are the bias compared to $M_E^*$, the empirical standard errors over replicates, the mean asymptotic nuisance-adjusted standard error, and the coverage for the GEE, the IPW, the AUG and the DR with independence (-I) and exchangeable (-E) working correlation matrix. }
\small
\begin{tabular}{rccccccccccccc}
  \hline
  &&\multicolumn{2}{c}{\textbf{ }} &\multicolumn{6}{c}{\textbf{Standard Error (SE)}} &\multicolumn{4}{c}{\textbf{Coverage}} \\
 &&\multicolumn{2}{c}{\textbf{Bias}} &\multicolumn{2}{c}{\textbf{Empirical }} &\multicolumn{2}{c}{\textbf{Robust}}  &\multicolumn{2}{c}{\textbf{Fay's}} &\multicolumn{2}{c}{\textbf{Robust}} &\multicolumn{2}{c}{\textbf{Fay's}}\\
 &$\boldsymbol{M_E^*}$ & -I & -E & -I & -E & -I & -E & -I & -E & -I & -E & -I & -E  \\
  \hline
     \multicolumn{14}{l}{\textbf{ Small sample $M=10, n_i=(10,20,30)$}with probability 1/3 each}\\
  ~~~~GEE &5.73&   2.214 & 2.213 & 1.330 & 1.329 & 1.829 & 1.848  & 1.359 & 1.363 & 89.4 & 88.7 & 59.4 & 59.7  \\ 
  ~~~~IPW &5.73&0.536 & 0.537 & 1.333 & 1.333 & 1.214 & 1.214 & 1.470 & 1.471  & 86.5 & 86.4 & 92.1 & 92.1  \\ 
    ~~~~AUG &5.73& 0.173 & 0.173 & 0.973 & 0.973  & 0.878 & 0.878 & 0.925 & 0.925 & 88.6 & 88.6 & 89.9 & 89.9  \\
  ~~~~DR &5.73&  -0.104 & -0.104 & 1.102 & 1.101 & 0.932 & 0.931 & 0.982 & 0.982 & 90.3 & 90.3 & 92.0 & 92.0  \\\\  
    \hline
     \multicolumn{14}{l}{\textbf{SAM-like sample $M=50, n_i=(20,30,30)$} with probability 1/3 each}\\
  ~~~~GEE &5.73& 2.347 & 2.343 & 0.308 & 0.308 & 0.532 & 0.466 & 0.308 & 0.309 & 0.0 & 0.0 & 0.0 & 0.0  \\ 
  ~~~~IPW &5.73& 0.622 & 0.623 & 0.303 & 0.303 & 0.317 & 0.317 & 0.323 & 0.323 & 50.7 & 50.7 & 52.1 & 52.1 \\ 
  ~~~~AUG &5.73& 0.215 & 0.215 & 0.222 & 0.222 & 0.230 & 0.230 & 0.232 & 0.232  & 85.1 & 85.1 & 85.2 & 85.2  \\ 
  ~~~~DR &5.73&  0.037 & 0.026 & 0.259 & 0.260 & 0.252 & 0.253 & 0.254 & 0.255 & 94.6 & 95.3 & 94.6 & 95.4   \\ 
 \hline
 \multicolumn{14}{l}{\textbf{Marginal model for the GEE:}}\\
  \multicolumn{14}{l}{$ \qquad \qquad  \mu(\boldsymbol \beta, A_i)= \beta_0+ \beta_A A_i$}\\  \multicolumn{14}{l}{\small \textbf{OM in AUG and DR is fitted for each treatment group using a stepwise regression:}}\\
    \multicolumn{14}{l}{$\qquad \qquad  B(\boldsymbol X_i, A_i=a)= \text{stepwise}({\rm EMP}_{ij},{\rm MARI}_{ij},{\rm AGE}_{ij},{\rm REL}_{ij},{\rm CAGE}_{ij},{\rm HIV}_{ij},{\rm CDM}_{ij},{\rm X1}_{ij},{\rm X2}_{ij},{\rm X3}_{ij}) $}\\
    \multicolumn{14}{l}{\small \textbf{PS in IPW and DR is fitted for the whole dataset using a stepwise regression:}}\\
    \multicolumn{14}{l}{$\qquad \qquad   logit(\pi_{ij}(\boldsymbol X_i, A_i))= \text{stepwise}({\rm A}_i,{\rm EMP}_{ij},{\rm MARI}_{ij},{\rm AGE}_{ij},{\rm REL}_{ij},{\rm CAGE}_{ij},{\rm HIV}_{ij},{\rm CDM}_{ij},{\rm X1}_{ij},{\rm X2}_{ij},{\rm X3}_{ij}) $}\\
     \hline
\end{tabular}
\label{tab:simulreel}
\end{table}

\section{Discussion}

We propose a doubly robust method for the estimation of the marginal effect of treatment in CRTs with continuous data subject to rMAR - an assumption that arises because missingness is  non-monotone in CRTs. Extension to binary or other outcomes is straightforward, provided that there is a one-to-one link function $h$ such that: $\mu_{ij}=h(\boldsymbol X_i,A_i)$. We extend the IPW approach proposed by \citet{robins1995analysis} and  the AUG approach for CRTs proposed by \citet{stephens2012augmented}. To be CAN, the DR estimator requires  that either the OM or PS model be correctly specified regardless of the choice of the working correlation matrix. Interfering covariates can be ignored  if either the OM or the PS is correctly specified. In presence of treatment-covariate interactions, if the PS is not correctly specified, covariates that interact with treatment on the outcome must be included in the OM. We accommodate these treatment-covariate interactions by modeling the OM separately for each treatment group. Covariates for the OM and the PS may be selected using automatic variable selection procedures such as a stepwise procedure, and may be at the cluster level or individual level. 

We recommend using $ \boldsymbol  V_i^{-1} \boldsymbol  W_i(\boldsymbol X_i, A_i, \boldsymbol \eta_W)$ \textcolor{black}{to ensure consistency of} the IPW and the DR for CRTs, rather than the conventional implementation, $\boldsymbol W_i^{1/2}(\boldsymbol \eta_W) \boldsymbol  V_i^{-1} \boldsymbol  W_i^{1/2}(\boldsymbol \eta_W)$, available in several software packages of the weighted GEE. See \citet{tchetgen2012weight} for a similar result for longitudinal data with observation-specific weights. If a working independence correlation structure is used, then the two implementations lead to the same result. When $\boldsymbol W_i^{1/2}(\boldsymbol \eta_W) \boldsymbol  V_i^{-1} \boldsymbol  W_i^{1/2}(\boldsymbol \eta_W)$ and an arbitrary correlation structure is used in the DR, estimation of marginal treatment effect is consistent  only if the OM is correctly specified.  We provide an R package called \textit{CRTgeeDR} that implements the proposed DR estimator. The application of our methods to data from the SAM study showed an effect of HIV/STI intervention on the percentage of protected intercourse  \citep{jemmott2014cluster} that reached a 0.05 level of significance. Moreover, results of the analysis that distinguishes among different types of partners and of sexual behavior  may be useful in targeting future interventions.
 Our approach allows a situation that we denoted covariate interference in CRTs, and thus extends ideas of adjustment of time-varying covariates in longitudinal responses \citep{sullivan1994cautionary,tchetgen2012weight}. Since treatment is randomized at a cluster level and we consider a marginal mean model which only includes treatment, the covariate interference have a different implication for analysis than exposure interference in causal framework \citep{liu2014large} or confounding by cluster in observational studies \citep{berlin1999empirical,huang2011informative}. However, when there are interactions between $ X^r_{ij}$ and $A_i$ exposure and  covariate interference are related; in this case, individual $ij$ may be seen as receiving pseudo-treatment $A_i  X^r_{ij}$. For such a setting, our work may be seen as extending the notion of exposure interference in RTs to CRTs and is  related to the work of \citet{ogburn2014causal}. In any case, modeling covariate interference may lead to substantial gains of efficiency if they predict the outcome. Therefore, it may be profitable to develop methods that make use of contact network information to inform the selection of interfering covariates. Finally, an IPW sensitivity analysis to address outcome MNAR as in \citet{rotnitzky1998semiparametric,vansteelandt2007estimation} would be useful to developed.

\vspace{-0.75cm}
\section{Web-Supplementary Materials}
Web Appendices, Tables, Figures, simulated data and, R sources implementing the estimators referenced in Sections \ref{part:supp1}, \ref{part:supp2} and \ref{part:mimick} are available with this paper at the Biometrics website on Wiley Online Library.

\vspace{-0.75cm}
\section*{Acknowledgements}
We thank J. Jemmot for sharing the SAM study  (NIH grant 1 R01 HD053270). This work was founded by NIH grants R37 AI 51164, AI 24643, AI113251, ES020337 and AI104459. 
Portions of this research were conducted on the Cluster at Harvard Medical (NIH grant NCRR 1S10RR028832-01).

\vspace{-1cm}
\bibliographystyle{biom.bst}
\bibliography{GEE_20150126.bib}

\begin{thebibliography}{}

\bibitem[\protect\citeauthoryear{Belitser, Martens, Pestman, Groenwold, Boer,
  and Klungel}{Belitser et~al.}{2011}]{belitser2011measuring}
Belitser, S.~V., Martens, E.~P., Pestman, W.~R., Groenwold, R.~H., Boer, A.,
  and Klungel, O.~H. (2011).
\newblock Measuring balance and model selection in propensity score methods.
\newblock {\em Pharmacoepidemiology and drug safety} {\bf 20,} 1115--1129.

\bibitem[\protect\citeauthoryear{Berlin, Kimmel, Have, and Sammel}{Berlin
  et~al.}{1999}]{berlin1999empirical}
Berlin, J.~A., Kimmel, S.~E., Have, T. R.~T., and Sammel, M.~D. (1999).
\newblock An empirical comparison of several clustered data approaches under
  confounding due to cluster effects in the analysis of complications of
  coronary angioplasty.
\newblock {\em Biometrics} {\bf 55,} 470--476.

\bibitem[\protect\citeauthoryear{Beunckens, Sotto, and Molenberghs}{Beunckens
  et~al.}{2008}]{beunckens2008simulation}
Beunckens, C., Sotto, C., and Molenberghs, G. (2008).
\newblock A simulation study comparing weighted estimating equations with
  multiple imputation based estimating equations for longitudinal binary data.
\newblock {\em Computational Statistics \& Data Analysis} {\bf 52,} 1533--1548.

\bibitem[\protect\citeauthoryear{Brookhart, Schneeweiss, Rothman, Glynn, Avorn,
  and St{\"u}rmer}{Brookhart et~al.}{2006}]{brookhart2006variable}
Brookhart, M.~A., Schneeweiss, S., Rothman, K.~J., Glynn, R.~J., Avorn, J., and
  St{\"u}rmer, T. (2006).
\newblock Variable selection for propensity score models.
\newblock {\em American journal of epidemiology} {\bf 163,} 1149--1156.

\bibitem[\protect\citeauthoryear{Brumback, Dailey, Brumback, Livingston, and
  He}{Brumback et~al.}{2010}]{brumback2010adjusting}
Brumback, B.~A., Dailey, A.~B., Brumback, L.~C., Livingston, M.~D., and He, Z.
  (2010).
\newblock Adjusting for confounding by cluster using generalized linear mixed
  models.
\newblock {\em Statistics \& probability letters} {\bf 80,} 1650--1654.

\bibitem[\protect\citeauthoryear{Brumback and He}{Brumback and
  He}{2011}]{brumback2011adjusting}
Brumback, B.~A. and He, Z. (2011).
\newblock Adjusting for confounding by neighborhood using complex survey data.
\newblock {\em Stat. Med.} {\bf 30,} 965--972.

\bibitem[\protect\citeauthoryear{Glynn and Quinn}{Glynn and
  Quinn}{2010}]{glynn2010introduction}
Glynn, A.~N. and Quinn, K.~M. (2010).
\newblock An introduction to the augmented inverse propensity weighted
  estimator.
\newblock {\em Political Analysis} {\bf 18,} 36--56.

\bibitem[\protect\citeauthoryear{Hawkes, Sivasivugha, Ngigi, Masumbuko, Brophy,
  and Kibendelwa}{Hawkes et~al.}{2013}]{hawkes2013hiv}
Hawkes, M., Sivasivugha, E.~S., Ngigi, S.~K., Masumbuko, C.~K., Brophy, J., and
  Kibendelwa, Z.~T. (2013).
\newblock \uppercase{HIV} and religion in the congo: A mixed-methods study.
\newblock {\em Current \uppercase{HIV} research} {\bf 11,} 246--253.

\bibitem[\protect\citeauthoryear{Huang and Leroux}{Huang and
  Leroux}{2011}]{huang2011informative}
Huang, Y. and Leroux, B. (2011).
\newblock Informative cluster sizes for subcluster-level covariates and
  weighted generalized estimating equations.
\newblock {\em Biometrics} {\bf 67,} 843--851.

\bibitem[\protect\citeauthoryear{Hubbard, Ahern, Fleischer, Van~der Laan,
  et~al\mbox{.}}{Hubbard et~al.}{2010}]{hubbard2010gee}
Hubbard, A.~E., Ahern, J., Fleischer, N.~L., Van~der Laan, M., et~al. (2010).
\newblock To \uppercase{GEE} or not to \uppercase{GEE}: comparing population
  average and mixed models for estimating the associations between neighborhood
  risk factors and health.
\newblock {\em Epidemiology} {\bf 21,} 467--474.

\bibitem[\protect\citeauthoryear{Hudgens and Halloran}{Hudgens and
  Halloran}{2012}]{hudgens2012toward}
Hudgens, M.~G. and Halloran, M.~E. (2012).
\newblock Toward causal inference with interference.
\newblock {\em JASA} {\bf 103,} 832--842.

\bibitem[\protect\citeauthoryear{Jemmott~III, Jemmott, O’Leary,
  et~al\mbox{.}}{Jemmott~III et~al.}{2014}]{jemmott2014cluster}
Jemmott~III, J.~B., Jemmott, L.~S., O’Leary, A., et~al. (2014).
\newblock Cluster-randomized controlled trial of an \uppercase{HIV}/sexually
  transmitted infection risk-reduction intervention for south african men.
\newblock {\em American journal of public health} {\bf 104,} 467--473.

\bibitem[\protect\citeauthoryear{Kaiser, Bunnell, Hightower,
  et~al\mbox{.}}{Kaiser et~al.}{2011}]{kaiser2011factors}
Kaiser, R., Bunnell, R., Hightower, A., et~al. (2011).
\newblock Factors associated with hiv infection in married or cohabitating
  couples in kenya: results from a nationally representative study.
\newblock {\em PLoS One} {\bf 6,} e17842.

\bibitem[\protect\citeauthoryear{Li, Shen, Li, and Robins}{Li
  et~al.}{2011}]{li2011weighting}
Li, L., Shen, C., Li, X., and Robins, J.~M. (2011).
\newblock On weighting approaches for missing data.
\newblock {\em Statistical methods in medical research} {\bf 22,} 14--30.

\bibitem[\protect\citeauthoryear{Liang and Zeger}{Liang and
  Zeger}{1986}]{liang1986longitudinal}
Liang, K.-Y. and Zeger, S.~L. (1986).
\newblock Longitudinal data analysis using generalized linear models.
\newblock {\em Biometrika} {\bf 73,} 13--22.

\bibitem[\protect\citeauthoryear{Liu and Hudgens}{Liu and
  Hudgens}{2014}]{liu2014large}
Liu, L. and Hudgens, M.~G. (2014).
\newblock Large sample randomization inference of causal effects in the
  presence of interference.
\newblock {\em J. Am. Stat. Asso.} {\bf 109,} 288--301.

\bibitem[\protect\citeauthoryear{McDaniel, Henderson, and Rathouz}{McDaniel
  et~al.}{2013}]{mcdaniel2013fast}
McDaniel, L.~S., Henderson, N.~C., and Rathouz, P.~J. (2013).
\newblock Fast pure r implementation of gee: Application of the matrix package.
\newblock {\em The R journal} {\bf 5,} 181.

\bibitem[\protect\citeauthoryear{Moore and van~der Laan}{Moore and van~der
  Laan}{2009}]{moore2009increasing}
Moore, K. and van~der Laan, M. (2009).
\newblock Increasing power in randomized trials with right censored outcomes
  through covariate adjustment.
\newblock {\em J. biopharm. stat.} {\bf 19,} 1099--1131.

\bibitem[\protect\citeauthoryear{Ogburn and VanderWeele}{Ogburn and
  VanderWeele}{2014}]{ogburn2014causal}
Ogburn, E.~L. and VanderWeele, T.~J. (2014).
\newblock Causal diagrams for interference.
\newblock {\em Statistical Science} {\bf 29,} 559--578.

\bibitem[\protect\citeauthoryear{Paik}{Paik}{1997}]{paik1997generalized}
Paik, M.~C. (1997).
\newblock The generalized estimating equation approach when data are not
  missing completely at random.
\newblock {\em JASA} {\bf 92,} 1320--1329.

\bibitem[\protect\citeauthoryear{Pepe and Anderson}{Pepe and
  Anderson}{1994}]{sullivan1994cautionary}
Pepe, M.~S. and Anderson, G.~L. (1994).
\newblock A cautionary note on inference for marginal regression models with
  longitudinal data and general correlated response data.
\newblock {\em Communications in Statistics-Simulation and Computation} {\bf
  23,} 939--951.

\bibitem[\protect\citeauthoryear{Robins}{Robins}{2000}]{robins2000marginal}
Robins, J.~M. (2000).
\newblock Marginal structural models versus structural nested models as tools
  for causal inference.
\newblock In {\em Statistical models in epidemiology, the environment, and
  clinical trials}, pages 95--133. Springer.

\bibitem[\protect\citeauthoryear{Robins, Rotnitzky, and Zhao}{Robins
  et~al.}{1994}]{robins1994estimation}
Robins, J.~M., Rotnitzky, A., and Zhao, L.~P. (1994).
\newblock Estimation of regression coefficients when some regressors are not
  always observed.
\newblock {\em JASA} {\bf 89,} 846--866.

\bibitem[\protect\citeauthoryear{Robins, Rotnitzky, and Zhao}{Robins
  et~al.}{1995}]{robins1995analysis}
Robins, J.~M., Rotnitzky, A., and Zhao, L.~P. (1995).
\newblock Analysis of semiparametric regression models for repeated outcomes in
  the presence of missing data.
\newblock {\em JASA} {\bf 90,} 106--121.

\bibitem[\protect\citeauthoryear{Rosenbaum}{Rosenbaum}{2007}]{rosenbaum2007interference}
Rosenbaum, P.~R. (2007).
\newblock Interference between units in randomized experiments.
\newblock {\em JASA} {\bf 102,} 191--200.

\bibitem[\protect\citeauthoryear{Rotnitzky, Robins, and Scharfstein}{Rotnitzky
  et~al.}{1998}]{rotnitzky1998semiparametric}
Rotnitzky, A., Robins, J.~M., and Scharfstein, D.~O. (1998).
\newblock Semiparametric regression for repeated outcomes with nonignorable
  nonresponse.
\newblock {\em JASA} {\bf 93,} 1321--1339.

\bibitem[\protect\citeauthoryear{Rubin}{Rubin}{1976}]{rubin1976inference}
Rubin, D.~B. (1976).
\newblock Inference and missing data.
\newblock {\em Biometrika} {\bf 63,} 581--592.

\bibitem[\protect\citeauthoryear{SAS}{SAS}{2015}]{SASGENMOD}
SAS (2015).
\newblock The genmod procedure sas 12.3.
\newblock {\em http://support.sas.com/documentation/} .

\bibitem[\protect\citeauthoryear{Seaman, Pavlou, and Copas}{Seaman
  et~al.}{2014}]{seaman2014review}
Seaman, S., Pavlou, M., and Copas, A. (2014).
\newblock Review of methods for handling confounding by cluster and informative
  cluster size in clustered data.
\newblock {\em Stat. Med.} {\bf 33,} 5371--5387.

\bibitem[\protect\citeauthoryear{Seaman and White}{Seaman and
  White}{2013}]{seaman2013review}
Seaman, S.~R. and White, I.~R. (2013).
\newblock Review of inverse probability weighting for dealing with missing
  data.
\newblock {\em Statistical Methods in Medical Research} {\bf 22,} 278--295.

\bibitem[\protect\citeauthoryear{Stephens, Tchetgen~Tchetgen, and
  Gruttola}{Stephens et~al.}{2012}]{stephens2012augmented}
Stephens, A.~J., Tchetgen~Tchetgen, E.~J., and Gruttola, V.~D. (2012).
\newblock Augmented generalized estimating equations for improving efficiency
  and validity of estimation in cluster randomized trials by leveraging
  cluster-level and individual-level covariates.
\newblock {\em Stat. Med.} {\bf 31,} 915--930.

\bibitem[\protect\citeauthoryear{Tchetgen~Tchetgen, Glymour, Weuve, and
  Shpitser}{Tchetgen~Tchetgen et~al.}{2012}]{tchetgen2012weight}
Tchetgen~Tchetgen, E., Glymour, M., Weuve, J., and Shpitser, I. (2012).
\newblock Specifying the correlation structure in inverse-probability-
  weighting estimation for repeated measures.
\newblock {\em Epidemiology} {\bf 23,} 644--646.

\bibitem[\protect\citeauthoryear{Tchetgen~Tchetgen and
  VanderWeele}{Tchetgen~Tchetgen and VanderWeele}{2012}]{tchetgen2012causal}
Tchetgen~Tchetgen, E.~J. and VanderWeele, T.~J. (2012).
\newblock On causal inference in the presence of interference.
\newblock {\em Statistical Methods in Medical Research} {\bf 21,} 55--75.

\bibitem[\protect\citeauthoryear{Tsiatis}{Tsiatis}{2006}]{tsiatis2006improving}
Tsiatis, A.~A. (2006).
\newblock Improving efficiency and double robustness with coarsened data.
\newblock {\em Semiparametric Theory and Missing Data} pages 221--272.

\bibitem[\protect\citeauthoryear{Tsiatis, Davidian, Zhang, and Lu}{Tsiatis
  et~al.}{2008}]{tsiatis2008covariate}
Tsiatis, A.~A., Davidian, M., Zhang, M., and Lu, X. (2008).
\newblock Covariate adjustment for two-sample treatment comparisons in
  randomized clinical trials: A principled yet flexible approach.
\newblock {\em Stat. Med.} {\bf 27,} 4658--4677.

\bibitem[\protect\citeauthoryear{Van~der Laan and Robins}{Van~der Laan and
  Robins}{2003}]{van2003unified}
Van~der Laan, M.~J. and Robins, J.~M. (2003).
\newblock {\em Unified methods for censored longitudinal data and causality}.
\newblock Springer Science \& Business Media.

\bibitem[\protect\citeauthoryear{Vansteelandt}{Vansteelandt}{2007}]{vansteelandt2007confounding}
Vansteelandt, S. (2007).
\newblock On confounding, prediction and efficiency in the analysis of
  longitudinal and cross-sectional clustered data.
\newblock {\em Scand. J. Stat.} {\bf 34,} 478--498.

\bibitem[\protect\citeauthoryear{Vansteelandt, Rotnitzky, and
  Robins}{Vansteelandt et~al.}{2007}]{vansteelandt2007estimation}
Vansteelandt, S., Rotnitzky, A., and Robins, J. (2007).
\newblock Estimation of regression models for the mean of repeated outcomes
  under nonignorable nonmonotone nonresponse.
\newblock {\em Biometrika} {\bf 94,} 841--860.

\bibitem[\protect\citeauthoryear{Zeger and Liang}{Zeger and
  Liang}{1986}]{zeger1986longitudinal}
Zeger, S.~L. and Liang, K.-Y. (1986).
\newblock Longitudinal data analysis for discrete and continuous outcomes.
\newblock {\em Biometrics} {\bf 42,} 121--130.

\bibitem[\protect\citeauthoryear{Zhang, Tsiatis, and Davidian}{Zhang
  et~al.}{2008}]{zhang2008improving}
Zhang, M., Tsiatis, A.~A., and Davidian, M. (2008).
\newblock Improving efficiency of inferences in randomized clinical trials
  using auxiliary covariates.
\newblock {\em Biometrics} {\bf 64,} 707--715.

\end{thebibliography}

\label{lastpage}

\end{document}